\documentclass[pra,twocolumn,longbibliography,nofootinbib]{revtex4-1}

\usepackage[utf8]{inputenc}
\usepackage{bm,amsfonts,amsmath,amssymb}
\usepackage{dcolumn,xfrac}
\usepackage{natbib}
\usepackage{tipa}
\usepackage{booktabs}
\usepackage{multirow}
\newcommand{\Eref}[1]{Eq.~(\ref{#1})}
\newcommand{\Tref}[1]{Table~\ref{#1}}
\newcommand{\Fref}[1]{Figure~\ref{#1}}

\def\vkapp{\varkappa}

\begin{document}
\title{Calculation of hyperfine magnetic anomaly in many-electron atoms}
\author{E. A. Konovalova$^{1}$}
\author{Yu. A. Demidov$^{1,2}$}
\author{M. G. Kozlov$^{1,2}$}
\affiliation{$^1$ Petersburg Nuclear Physics Institute of NRC ``Kurchatov center'', 188300 Gatchina, Leningrad District, Russia}
\affiliation{$^2$ St.~Petersburg Electrotechnical University ``LETI'', Prof. Popov Str. 5, 197376 St. Petersburg, Russia}
\date{\today}
\begin{abstract}
The precision measurements of the ratio of hyperfine structure constants for $s_{1/2} $ and $ p_{1/2} $ states
allow us to estimate the difference between hyperfine magnetic anomalies for these levels.
We calculate the atomic factor in order to recover the absolute values of the hyperfine magnetic anomalies from their difference.
Taking into account the hyperfine anomaly  correction allows one to increase the accuracy of determining the $g$ factors of short-lived isotopes by more than an order of magnitude.\\[3mm]
\end{abstract}
\maketitle
\section{Introduction}
The accuracy achieved in recent years in laser spectroscopy coupled with the advanced atomic theory allowed one 
to test various nuclear models~\cite{Z2015,G1999}. 
The magnetic hyperfine structure (HFS) constants $A$ depend on the nuclear charge and nuclear magnetization distributions. 
For the point-like nucleus the ratio of hyperfine constants $A$ for different isotopes is equal to the ratio of their nuclear $g$ factors $g_I = \frac{\mu}{\mu_N I}$,
where $\mu$ and $I$ are nuclear magnetic moment and nuclear spin, and $\mu_N$ is the nuclear magneton.
Taking into account finite nuclear size we have to consider the effect produced on the hyperfine constants by
the nuclear magnetization distribution and the behavior of the electronic wave function inside the nucleus.
Former correction is called magnetic, or Bohr--Weisskopf (BW) correction~\cite{BW50}), and the latter one is called charge,
or Breit-Rosenthal (BR) correction~\cite{RB32,CS49}.
These corrections  break proportionality between the HFS constants and the nuclear $g$ factors.
 This phenomenon is called the hyperfine anomaly (HFA)~\cite{BW50}. For the isotopes (1) and (2) the HFA is defined as:
 \begin{align}
 \label{hfs_anomaly}
 ^1\Delta^2 \equiv \frac{A^{(1)}g_I^{(2)}}{A^{(2)}g_I^{(1)}} -1.
  \end{align}

Usually nuclear $g$ factors of short-lived isotopes are extracted from the measured HFS constants neglecting the HFA correction.
On the one hand, the necessary theoretical results are not available. On the other hand, the HFA correction is usually small (less than 1\%)~\cite {Per13}. Only recently the measurements for short-lived isotopes reached this level of accuracy. Note that an accurate study of the HFS constants can serve as a useful tool for understanding the phenomenon of shape coexistence in atomic nuclei~\cite {shape_coh}. Thus, the development of new theoretical methods for the HFS constants taking into account the HFA becomes relevant.

\section{Finite nuclear size corrections to HFS constants}\label{Sec_Theory}
It is generally accepted that the magnetic hyperfine structure constant $A$ can be written in the following form~\cite{stroke61}:
 \begin{align}
 \label{hfs_Shabaev_1}
 A = g_I {\cal A}_0 (1-\delta) (1-\epsilon).
 \end{align}
Here $g_I$ is nuclear $g$ factor, $g_I{\cal A}_0$ is the HFS constant for point-like nucleus, and
$\delta$ and $\epsilon$ are dimensionless BE and BW correction, respectively. 
Note that parameter ${\cal A}_0$ is independent on the nuclear $g$ factor. 
 
An analytical expression for the parameter ${\cal A}_0$ for the hydrogen-like ions is known from Refs.~\cite{pyykko73,Sha94}: 
 \begin{align}\label{hfs_Shabaev_2}
 {\cal A}_0 = \frac{\alpha (\alpha Z)^3}{j(j+1)}\,
 \frac{m}{m_p}\, \frac{\vkapp(2\vkapp(n_r+\gamma) - N)}{N^4\gamma(4\gamma^2 -1)}\, mc^2.
 \end{align}
Here $\alpha$ is fine structure constant, $Z$ is nuclear charge, $m$ and $m_p$ are the electron and proton masses, $j$ is total electron angular moment, 
$\vkapp =(l-j)(2j+1)$ is relativistic quantum number, $N= \sqrt{ (n_{r} +\gamma)^2 + (\alpha Z)^2}$ is effective principal quantum number, 
$n_r = n - |\vkapp|$ is radial quantum number, $n$ is nonrelativistic quantum number, and $\gamma = \sqrt{\vkapp^2 - (\alpha Z)^2}$.
Note, that for the levels with $j = \frac{1}{2}$ the parameter ${\cal A}_0$ goes to infinity when $\gamma  \to \frac{1}{2}$, which happens at $Z \approx 118$.

When the quantum numbers $n$ are the same, the ratios ${\cal A}_{0,\, s_{1/2}}/{\cal A}_{0,\, p_{1/2}}$ and ${\cal A}_{0,\, p_{1/2}}/{\cal A}_{0,\, p_{3/2}}$ are equal to:
 \begin{align}
 \label{A0_ratio}
 \begin{split}
 \frac{{\cal A}_{0,\, s_{1/2}}}{{\cal A}_{0,\, p_{1/2}}} &= 3+ 2\left( \frac{\alpha Z}{n -1 +\gamma_{1/2}} \right)^2;\\
 \frac{{\cal A}_{0,\, p_{1/2}}}{{\cal A}_{0,\, p_{3/2}}} &= \frac{\gamma_{3/2}(4\gamma^2_{3/2}-1)}{2\gamma_{1/2}(4\gamma^2_{1/2}-1)}\left(1 +3(\alpha Z)^2\frac{5n-4}{20n^2} \right).
 \end{split}
 \end{align}
Here we took into account that $\gamma$ is the same for $s_{1/2}$ and $p_{1/2}$ states, but is different for $p_{1/2}$ and $p_{3/2}$ states.
Equations \eqref{A0_ratio} are consistent with the results obtained for many electron systems in the semiclassical approximation~\cite{SFK78}.

In the following we use model of the homogeneously charged and magnetized ball of the radius $R = \sqrt{5/3} r_\mathrm{rms}$,
where \mbox{$r_\mathrm{rms}=\langle r^2\rangle^{1/2}$} is root-mean-square nuclear radius.
The nuclear magnetization distribution is caused by the spin polarization of the nucleons and the orbital motion of the protons.

The~charge density inside the nucleus is relatively stable for different isotopes~\cite{fermi_model},
whereas the nuclear magnetization strongly depends on the spin and configuration of each isotope.
Following Refs.\ \cite{BW50,B51,odd_odd_fr,MP95}, we introduce the nuclear factor $d_\mathrm{nuc}$ for parameterization of these nuclear effects.
Then, the BR and BW corrections $\delta$ and $\epsilon$ for a given $Z$ and given electron state can be written as \cite{KKDB17}:
 \begin{align}
 \label{H-scalings}
 \begin{split}
\delta(R) &= b_N (R/\lambdabar_C)^{2\gamma -1}, \\
\epsilon(R, d_\mathrm{nuc}) &= b_M d_\mathrm{nuc}(R/\lambdabar_C )^{2\gamma -1}.
\end{split}
\end{align}
Here $b_N$ and $b_M$ are dimensionless factors, which are independent of the nuclear radius and structure;
$\lambdabar_C$ is the electron Compton wavelength \mbox{($\lambdabar_C =\frac{\hbar}{m_e c}$)}.
The nuclear factor is defined so that $d_\mathrm{nuc}=0$ corresponds to a point-like magnetic dipole in the center of the nucleus and 
$d_\mathrm{nuc}=1$ to corresponds to the homogeneously magnetized ball of radius $R$.

The HFS parametrization Eqs.~(\ref{hfs_Shabaev_1}, \ref{H-scalings}) includes three nuclear parameters, $g_I$, $d_\mathrm{nuc}$, and $R$ and three atomic ones, ${\cal A}_0$, $b_N$, and $b_M$. If we can calculate atomic parameters accurately enough, we can use experimental values of the HFS constants $A$ to get information about the nucleus. Atomic parameters are the same for different isotopes. Moreover, the charge radius $R$ typically differs by few percent, so the BR correction for all isotopes is almost the same. Therefore, the most important correction for the isotopic studies comes from the BW correction \eqref{H-scalings}. Consequently, the most important atomic parameter for these studies is $b_M$.

Before doing atomic calculations of the HFS constants $A$ we need to specify three nuclear parameters:
\begin{align}
\label{HFS_nuc}
\begin{split}
    A(g_I,d_\mathrm{nuc},R) =&g_I{\cal A}_0 \left(1 - b_{N} (R/\lambdabar_C )^{2\gamma -1}\right)\times\\
    &\times\left(1 - d_\mathrm{nuc} b_M (R/\lambdabar_C)^{2\gamma -1}\right).
\end{split}
\end{align}
In order to calculate atomic parameters ${\cal A}_0$, $b_N$, and $b_M$ from Eqs.\ (\ref{hfs_Shabaev_1}, \ref{H-scalings}),  we vary $d_\mathrm{nuc}$ and $R$, assuming $g_I = 1$. After that we can predict HFS constants for any nuclear parameters.

To find parameter $b_M$ we do calculation for the point-like magnetic dipole ($d_\mathrm{nuc}=0$) and for the homogeneously magnetized ball ($d_\mathrm{nuc}=1$). Then $b_M$ is given by:  
\begin{align}\label{b_M}
	b_M &= (R/\lambdabar_C)^{1-2\gamma} 
   \left(1 - 
    \frac{A(g_I,1,R)}{A(g_I,0,R)}
    \right)\,.
\end{align}
To find parameter $b_N$ we need calculations for different radii $R$: 
\begin{align}\label{b_N}
    b_N &= 
	\frac{\left(A(g_I,0,R_2) 
	- A(g_I,0,R_1)\right)
   \lambdabar_C^{2\gamma-1}}   
   {A(g_I,0,R_2)R_1^{2\gamma-1}
   -A(g_I,0,R_1)R_2^{2\gamma-1}}
   \,.
\end{align}
The remaining atomic parameter ${\cal A}_0$ is now found from the relation:
\begin{align}\label{A_0}
    {\cal A}_0 &= 
    \frac{A(g_I,0,R)}
	{g_I\left(1 - b_N (R/\lambdabar_C)^{2\gamma-1}\right)} 
    \,.
\end{align}

\section{Hyperfine magnetic anomaly}
Let us compare HFS constants for two isotopes with nuclear $g$ factors $g_I^{(1)}$ and $g_I^{(2)}$ and
slightly different nuclear radii $R^{(1,2)}=R\pm \mathfrak{r}$ to find the hyperfine magnetic anomaly $^1\Delta^2$ using \Eref{hfs_anomaly}.
The HFA can be divided in two terms associated with changes in charge $^1\Delta^2_\mathrm{BR}$ and magnetization $^1\Delta^2_\mathrm{BW}$ distributions:
$^1\Delta^2={}^1\Delta^2_\mathrm{BR}+{}^1\Delta^2_\mathrm{BW}$.
Assuming the nuclear factors of both isotopes $d_\mathrm{nuc}^{(1)}=d_\mathrm{nuc}^{(2)}=0$ we obtain:
\begin{align}
\frac{A(g_I^{(1)}, 0, R+\mathfrak{r})}{A(g_I^{(2)}, 0, R-\mathfrak{r})}
\approx \frac{g_I^{(1)}}{g_I^{(2)}} + 2\mathfrak{r}\frac{\partial A(g_I^{(1)}, 0, R)/\partial R}{A(g_I^{(2)}, 0, R)}.
\end{align}
Then the term $^1\Delta^2_\mathrm{BR}(R)$ associated with changes in charge distribution is equal to:
 \begin{align}\label{BR_anom}
 \begin{split}
    ^1\Delta^2_\mathrm{BR}(R,\mathfrak{r}) &\equiv \frac{g_I^{(2)}A(g_I^{(1)}, 0, R+\mathfrak{r})}{g_I^{(1)}A(g_I^{(2)}, 0, R-\mathfrak{r})}-1 \\
    &\approx -2(2\gamma -1) b_N \frac{R^{2\gamma -2}\mathfrak{r}}{\lambdabar_C^{2\gamma -1}}.
\end{split}
\end{align}
In the case when the nuclear factors for both isotopes are the same $d_\mathrm{nuc}^{(1)}=d_\mathrm{nuc}^{(2)}=d_\mathrm{nuc}$ (isotopes with identical spins and similar nuclear configurations) a similar expression can be obtained for HFA $^1\Delta^2$:
 \begin{align}\label{all_anom}
  \begin{split}
    ^1\Delta^2 &\equiv \frac{A(1, d_\mathrm{nuc}, R+\mathfrak{r})}{A(1, d_\mathrm{nuc}, R-\mathfrak{r})}-1 \approx\\
    &\approx -2(2\gamma -1) (b_N + d_\mathrm{nuc}b_M) \frac{R^{2\gamma -2}\mathfrak{r}}{\lambdabar_C^{2\gamma -1}}.
 \end{split}
\end{align}
However, the nuclear factors can vary significantly for isotopes with different nuclear spins.
In the case $d_\mathrm{nuc}^{(1)} \neq d_\mathrm{nuc}^{(2)}$ the difference between the  nuclear radii of considered isotopes can be neglected  and the HFA is given by the expression: 
 \begin{align}\label{dnuc_anom}
 ^1\Delta^2 \approx \left(d_\mathrm{nuc}^{(2)}-d_\mathrm{nuc}^{(1)}\right)\,b_M\left(\frac{R}{\lambdabar_C}\right )^{2\gamma -1} 
 = \epsilon_2-\epsilon_1.
\end{align}

\subsection{Nuclear factor}\label{Sec_dnuc}
Let us briefly describe $d_\mathrm{nuc}$ in terms of single-particle nuclear model.
In Refs.~\cite{BW50,B51} it is shown, that BW correction can be written as
 \begin{align}
 \label{BW}
         \epsilon = b_M (R/\lambdabar_C)^{2\gamma -1} \left( \left( 1+ \frac{2}{5} \zeta \right)\alpha_S + \frac{3}{5}\alpha_L \right) \frac{R_M^2}{R^2}.
 \end{align}
In this expression $\zeta$ is the spin asymmetry parameter:
\begin{align}
 \label{zeta}
\zeta = 
 \begin{cases}
	\frac{2I-1}{4(I+1)}, & \text{if $I = l+\frac{1}{2}$}, \\ 
	 \frac{2I+3}{4I}, & \text{if $I = l-\frac{1}{2}$}. \\ 
 \end{cases}
\end{align}
Coefficients $\alpha_S$ and $\alpha_L$ parametrize the spin $g_S$ and orbital $g_L$ contributions to the nuclear $g$ factor:
\begin{align}
 \label{alphas}
        &\alpha_S = \frac{g_S}{g_I} \frac{g_I - g_L}{g_S- g_L} \qquad \text{and} \qquad \alpha_L = 1-\alpha_S.
\end{align}
Finally, $R_M$ in \Eref{BW} is the radius of the nuclear magnetization density distribution.
The resulting expression for the nuclear factor $d_\mathrm{nuc}$ is easily seen to be
 \begin{align}
 \label{d_nuc}
         d_\mathrm{nuc} =  \left( \left( 1+ \frac{2}{5} \zeta \right)\alpha_S + \frac{3}{5}\alpha_L \right) \frac{R_M^2}{R^2}.
 \end{align}
The nuclear factor depends on the configuration of nucleons and may significantly vary from one isotope to another.
At the same time $d_\mathrm{nuc}$ weakly depends on the nuclear charge $Z$. For instance, if nuclear configuration includes single valence proton in the $s_{1/2}$ state, then $I = 1/2$, $\zeta = 0$, $\alpha_s = 1$, and, assuming $R_M\approx R$, we obtain $d_\mathrm{nuc} \approx 1$.
This case is realized, for example, for the stable isotopes of thallium.

For alkali metal atoms, the contributions of electronic correlations to the HFS constants can be taken into account very accurately (see e.g.~\cite{SJD99,GAOS08,sahoo15,GVF17}). Then the uncertainty in the nuclear factor becomes the major source of the theoretical error.
Now we will illustrate the uncertainty in the value of nuclear factor using the isotope $\rm ^{211}Fr$ as an example. According to the nuclear shell model this nucleus has single valence proton in the state $h_{9/2}$~\cite{G1999,Z2015}.
\begin{enumerate}
\item Using Eqs. (\ref{zeta} -- \ref{d_nuc}) we determine $\zeta = \frac{2}{3}$, $\alpha_S = -0.152$, $\alpha_L = 1.152$. Assuming that $R_M\approx R$, $g_L = 1$, and $g_S = g_{S, \,\mathrm{free}} = 5.586$ we get $d_\mathrm{nuc} = 0.50$.
For $\rm ^{209}Bi$ with the same nucleons configuration ($h_{9/2}$ proton state) the nuclear factor equal 0.47~\cite{Sha94}.
\item Following prescription of \citet{G1999}, M{\aa}rtesson-Pendrill took $g_L = 1.16$ and $g_S$ = 0.85$g_{S, \,\mathrm{free}}$ and deduced $d_\mathrm{nuc} = 0.33$ \cite{pendrill_fr}.
\item The nuclear factor can be deduced as a ratio of the Bohr-Weisskopf corrections obtained within the single-particle nuclear model and the model of the homogeneously magnetized ball. In the latter model model the nuclear factor is equal to 1.0. This way one gets $d_\mathrm{nuc}=0.54(21)$~\cite{GVF17}.
\item Using the radii $R = 7.281(65)$~fm~\cite{JS85} and $R_M = 6.71$~fm~\cite{SJD99} one can obtain nuclear factor as $d_\mathrm{nuc} =(R_M/R)^2=0.85$.
\end{enumerate}
These examples show that the nuclear factor and the Bohr--Weisskopf correction strongly depend on the nuclear model (see also~\cite{shab01,karpeshin15}).

\subsection{Differential hyperfine magnetic anomaly}\label{Sec_dhfa}

For a number of isotopic sequences the ratios of the HFS constants for low-lying $s_{1/2}$ and $p_{1/2}$ atomic states \mbox{$\rho = A(s_{1/2})/A(p_{1/2})$}
were measured with sufficient accuracy to reliably determine the differential HFS anomaly (DHFA):
\begin{align}\label{dhfa}
        {}^{\,\,\,1}_{s_{1/2}}\Delta^{2}_{p_{1/2}} = \frac{\rho^{(1)}}{\rho^{(2)}}-1\approx {}^1\Delta^2 (s_{1/2}) - {}^1\Delta^2(p_{1/2}) .
\end{align}

Differential hyperfine anomalies between isotopes with the same valence nuclear configuration are extremely small (about $10^{-4}$ even for heavy atoms~\cite{Per13}).
Below We will neglect these small changes of DHFA. Thus, we neglect the smooth changes of BR and BW corrections and restrict our attention to there-distributions of nuclear magnetization (i.e.\ changes of nuclear factor) in the case of isotopes with different nuclear spins. 
Then the HFA is given by \Eref{dnuc_anom} and the ratio of the hyperfine magnetic anomalies is determined solely by the following atomic factor (see e.g.~\cite{shab_prl06,schmidt18}):
\begin{align}\label{ratio}
        \eta = \frac{{}^1\Delta^2 (s_{1/2})}{{}^1\Delta^2 (p_{1/2})}
        \approx
        \frac{b_M(s_{1/2})}{b_M(p_{1/2})}\,.
\end{align}

Using the computed value of $\eta$ and assuming that $g$ factor $g_I^{(1)}$ is known, we can restore the hyperfine magnetic anomalies for $s_{1/2}$ and $p_{1/2}$ states and find the HFA correction to $g_I^{(2)}$:
\begin{align}\label{mu_corr}
\begin{split}
       {}^{1}\Delta^{2}(s_{1/2})&= \frac{{}^{\,\,\,1}_{s_{1/2}}\Delta^{2}_{p_{1/2}}}{1-1/\eta} ,\\
        g_I^{(2)}&= g_I^{(1)}\frac{A^{(2)}}{A^{(1)}}\left (1- {}^{1}\Delta^{2}(s_{1/2}) \right).
\end{split}
\end{align}
We can also determine the nuclear factor $d_\mathrm{nuc}^{(2)}$ in terms of $d_\mathrm{nuc}^{(1)}$ and the differential hyperfine magnetic anomaly:
\begin{align}\label{dhfa2}
	d_\mathrm{nuc}^{(2)} = d_\mathrm{nuc}^{(1)} + \frac{{}^{\,\,\,1}_{s_{1/2}}\Delta^{2}_{p_{1/2}}}{(1- 1/\eta)b_M(s_{1/2}) (R/\lambdabar_C)^{2\gamma -1}}.
\end{align}
\section{Results and discussion}
In this section we discuss the generic properties of HFS anomaly. In particular, we focus on the dependence of the Bohr--Weisskopf correction on the radius of the nuclear magnetization density distribution $R_M$ and the dependence of the atomic factor $\eta$ on $Z$. We start with the hydrogen-like ions, because for them there are analytical expressions and numerical calculations can be easily done for a wide range of atomic numbers. Then we show that for neutral atoms a similar analytical expression for $\eta$ can be obtained within one-particle approximation.  
The results of correlation calculations of the atomic factor for monovalent atoms, known in the literature, are consistent with the values obtained analytically.
All numerical calculations were performed with the program package \cite{KPST15}, which is based on the Hartree-Fock-Dirac code \cite{BDT77}.

\begin{figure}[h]
\centering
\includegraphics[height=8.5cm]{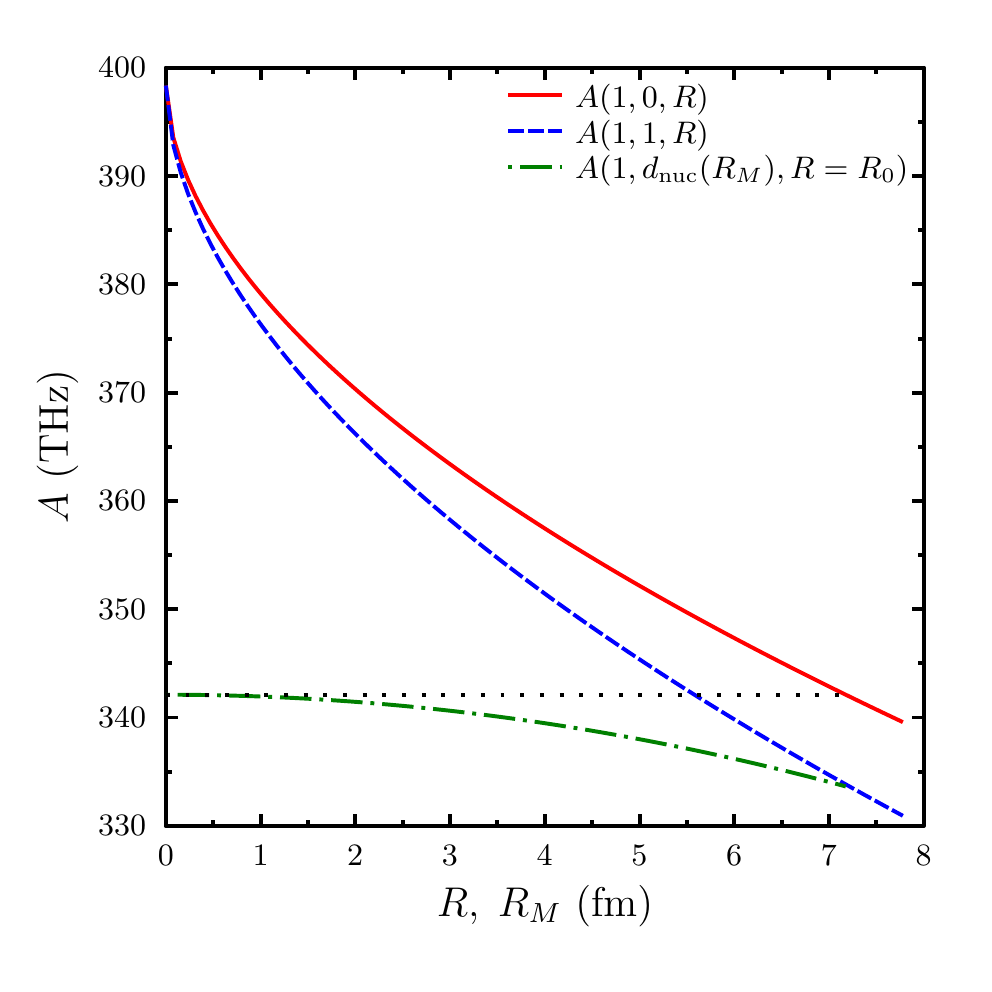}
\caption{Dependence of the HFS constant $A(g_I, d_\mathrm{nuc}(R_M), R)$ for the ground state of the H-like Fr ion on the nuclear radius $R$ and the radius of the nuclear magnetization density distribution $R_M$. We assume that $g_I =1$. Solid and dashed lines correspond to the fits by \Eref{HFS_nuc} when $d_\mathrm{nuc}$ is equal to 0 and 1 respectively. Dash dotted line corresponds to the case when $R =R_0$ is fixed and $R_M \in [0,\,R_0]$ changes ($d_\mathrm{nuc} \in [0,\,1]$). As follows from \Eref{d_nuc}, the vertex of the parabola lies on the horizontal dotted line, which crosses solid line at $R_M=R_0$. The parameter $R_0 = 7.214(23)$ fm is the nuclear radius of $\rm ^{211}Fr$~\cite{angeli13}.
 \label{frg:A_fit}}
\end{figure}

\subsection{Hydrogen-like ions}
\Fref{frg:A_fit} shows the dependence of the constant $A/g_I$ for the ground state of $\rm Fr^{86+}$ on the nuclear radius $R$
and the radius of the nuclear magnetization density distribution $R_M$.
Our numerical result for the HFS constant of the point nucleus ${\cal A}_0 (1s_{1/2})=398.4(6)$~THz is in good agreement with the analytical value  397.7~THz \cite{Sha94}. We do not take QED corrections into account which may be rather large for such a heavy system.
Equation \eqref{HFS_nuc} with $b_M(1s_{1/2})= 0.208(3)$ and $b_N(1s_{1/2}) = 1.24(2)$, found from \Eref{b_M} and \Eref{b_N} respectively, accurately describes the dependence of the HFS constant on the nuclear radius $R$ for $d_\mathrm{nuc}=0$, or 1.
If we fix $R=R_0$ and go from $R_M=R_0$  to $R_M=0$, then according to \Eref{d_nuc} the nuclear factor $d_\mathrm{nuc}(R_M)$ and BW correction change from their maximum values to zero proportionally to $R_M^2$.
 
\begin{table}[!htbp]
\caption{\label{tbl:BW_compare}
The dependence of the BW corrections $\epsilon (d_\mathrm{nuc},R)$~Eqs.~(\ref{H-scalings}, \ref{b_M}) for $2s_{1/2}$ and $2p_{1/2}$ states of H-like ions and their ratio $\eta = \frac{\epsilon(s_{1/2})}{\epsilon(p_{1/2})}$ from the nuclear charge  $Z$, assuming $d_\mathrm{nuc}=1$. 
Nuclear radii $R$ are taken from Ref.~\cite{angeli13}.}
        \begin{tabular}{lcccccccr}
\hline
\\[-3mm]
                Z&\multicolumn{3}{c}{$\epsilon (1, R) (2s_{1/2})$ (\%)}
                &\multicolumn{3}{c}{$\epsilon (1, R) (2p_{1/2})$ (\%)}
                &\multicolumn{2}{c}{$\eta$}\\
                &&\cite{Sha94}&\cite{BW50}&&\cite{Sha94}&\cite{BW50}&&\cite{Sha94}\\
\hline
                10 & 0.042  &0.043 & 0.05 & --  &0.0002& --  & --   &277.3\\
                20 & 0.106  &0.106 & 0.12 & --   &0.0015& --  & --  &73.9\\
                30 & 0.203  &0.205 & 0.23 &0.006 &0.006& 0.01&33.1  &33.2\\
                40 & 0.341  &0.344 & 0.41 &0.019 &0.019& 0.03&18.4&18.5\\
                50 & 0.553  &0.561 & 0.67 &0.048 &0.048& 0.08&11.5&11.6\\
                60 & 0.856  &0.873 & 1.03 &0.111 &0.112& 0.17&7.74&7.79\\
                70 & 1.335  &1.353 & 1.51 &0.245 &0.245& 0.36&5.45&5.52\\
                80 & 1.976      &2.048     & 2.15     &0.495     &0.505    & 0.70    &3.99&4.05\\
                {90~} & {~2.969~}  &{~3.077~} & {~2.88~} &{~0.987~} &{~1.004~}& {~1.27~}&{~3.01~}&{3.07}\\
\hline
\end{tabular}
\end{table}
\begin{table}[!htbp]
\caption{\label{tbl:BR_compare}
The dependence of the BR corrections $\delta (R)$~Eqs.~(\ref{H-scalings}, \ref{b_N}) for $2s_{1/2}$ and $2p_{1/2}$ states of H-like ions, their ratio \mbox{$\eta_{BR} = \frac{\delta(s_{1/2})}{\delta(p_{1/2})}$}, and coefficient $k = 1- \delta (R) (2s_{1/2})+\delta (R) (2p_{1/2})$ on the nuclear charge  $Z$. Nuclear radii $R$ are taken from Ref.~\cite{angeli13}.}
        \begin{tabular}{lddllrrc}
\hline
\\[-3mm]
                Z&\multicolumn{2}{c}{$\delta (R) (2s_{1/2})$ (\%)}
                &\multicolumn{2}{c}{$\delta (R) (2p_{1/2})$ (\%)}
                &\multicolumn{2}{c}{$\eta_{BR}$}&k\\
                &&\multicolumn{1}{c}{\cite{Sha94}}&&\multicolumn{1}{c}{\cite{Sha94}}&&\multicolumn{1}{c}{\cite{Sha94}}&\\
\hline
                10 & 0.115 &0.115 &\multicolumn{1}{c}{--}     
                                            & 0.0003&\multicolumn{1}{c}{--}
                                                          &383.3&1.00\\
                20 & 0.299 &0.295 & 0.0036  & 0.0036&82.6 & 81.9&1.00\\
                30 & 0.601 &0.593 & 0.0167  & 0.0165&36.0 & 35.9&0.99\\
                40 & 1.08  &1.06  &  0.0545 & 0.0539&19.8 & 19.7&0.99\\
                50 & 1.90  &1.86  &  0.156  & 0.154 &12.2 & 12.1&0.98\\
                60 & 3.28  &3.20  &  0.406  & 0.400 & 8.1 &  8.0&0.97\\
                70 & 5.83  &5.60  &  1.05   & 1.01  & 5.6 &  5.5&0.95\\
                80 & 10.1  &9.87  &  2.56   & 2.52  & 3.9 &  3.9&0.92\\
                90 & 18.3  &17.7  & 6.50    & 6.30  & 2.8 &  2.8&0.87\\
\hline
\end{tabular}
\end{table}
\begin{table*}[htbp]
\caption{\label{tbl:A0_compare}
The dependence of parameters ${\cal A}_0$ \Eref{A_0} for $2s_{1/2}$, $2p_{1/2}$ $2p_{3/2}$ states of H-like ions in comparison with analytical values~\Eref{hfs_Shabaev_2}~\cite{Sha94}. Comparison of numerical and analytical~\Eref{A0_ratio} values of $\frac{{\cal A}_{0,\, s_{1/2}}}{{\cal A}_{0,\, p_{1/2}}}$ and $\frac{{\cal A}_{0,\, p_{1/2}}}{{\cal A}_{0,\, p_{3/2}}}$ is presented in the last four columns.}
\centering
        \begin{tabular}{ldddddddddd}
\hline
\\[-3mm]
                Z&\multicolumn{2}{c}{${\cal A}_{0,\, s_{1/2}}$ (THz)}
                &\multicolumn{2}{c}{${\cal A}_{0,\, p_{1/2}}$ (THz)}
                &\multicolumn{2}{c}{${\cal A}_{0,\, p_{3/2}}$ (THz)}
                &\multicolumn{2}{c}{$\frac{{\cal A}_{0,\, s_{1/2}}}{{\cal A}_{0,\, p_{1/2}}}$} 
                &\multicolumn{2}{c}{$\frac{{\cal A}_{0,\, p_{1/2}}}{{\cal A}_{0,\, p_{3/2}}}$}\\
                Eq.&\multicolumn{1}{c}{\eqref{A_0}}&\multicolumn{1}{c}{\eqref{hfs_Shabaev_2}}
                &\multicolumn{1}{c}{\eqref{A_0}}&\multicolumn{1}{c}{\eqref{hfs_Shabaev_2}}
                &\multicolumn{1}{c}{\eqref{A_0}}&\multicolumn{1}{c}{\eqref{hfs_Shabaev_2}}
                &\multicolumn{1}{c}{\eqref{A_0}}&\multicolumn{1}{c}{\eqref{A0_ratio}}
                &\multicolumn{1}{c}{\eqref{A_0}}&\multicolumn{1}{c}{\eqref{A0_ratio}}\\
\hline
                10 & 0.032  &0.032 &0.0107 &0.0107&0.002&0.002  &3.00&3.00&5.04&5.04\\
                20 & 0.266  &0.266 & 0.088 &0.088 &0.017& 0.017 &3.01&3.01&5.18&5.18\\
                30 & 0.954  &0.954 & 0.315 &0.315 &0.058& 0.058 &3.02&3.02&5.43&5.43\\
                40 & 2.464  &2.463 & 0.809 &0.809 &0.139& 0.139 &3.05&3.04&5.81&5.81\\
                50 & 5.409  &5.408 & 1.761 &1.760 &0.275& 0.276 &3.07&3.07&6.39&6.38\\
                60 & 10.909 &10.905& 3.509 &3.509 &0.484& 0.485 &3.11&3.11&7.23&7.22\\
                70 & 21.200 &21.185& 6.718 &6.717 &0.785& 0.787 &3.16&3.15&8.54&8.49\\
                80 & 41.313 &41.261& 12.844 &12.841&1.202& 1.205&3.22&3.21&10.66&10.57\\
                90 & 84.547 &84.369& 25.660 &25.637&1.761& 1.767&3.29&3.28&14.51&14.31\\
\hline
\end{tabular}
\end{table*}

Tables \ref{tbl:BW_compare} and \ref{tbl:BR_compare} summarize numerical results for Bohr--Weisskopf and Breit--Rosenthal corrections
of $2s_{1/2}$ and $2p_{1/2}$ states of H-like ions for  $Z$ ranging  from 10 to 90. Computing BW corrections we assume $d_\mathrm{nuc} = 1$. 
Our numerical results for $\epsilon$ and $\delta$, as well, as their ratios $\eta = \frac{\epsilon(s_{1/2})}{\epsilon(p_{1/2})}$ and 
$\eta_\mathrm{BR} = \frac{\delta(s_{1/2})}{\delta(p_{1/2})}$ are in agreement with the analytical results from Ref.~\cite{Sha94}.
Note that both parameters $\eta$ and $\eta_\mathrm{BR}$ rapidly decrease with $Z$.

The dependence of parameters ${\cal A}_0$ for $2s_{1/2}$, $2p_{1/2}$, and $2p_{3/2}$ states of H-like ions from the nuclear charge is presented in \Tref{tbl:A0_compare}.
The wave function of the $p_{3/2}$ states turns to zero for $r=0$, thus the BW and BR corrections for these states are equal to zero. 
In this case calculated HFS constants for $g_I=1$ are equal to ${\cal A}_0$.
For $2s_{1/2}$ and $2p_{1/2}$ states the parameters ${\cal A}_0$ are obtained from \Eref{A_0}.
The parameters ${\cal A}_0$  differ from the analytical values \Eref{hfs_Shabaev_2} by less than 0.2\%, even for large $Z$.
The ratios $\frac{{\cal A}_{0,\, s_{1/2}}}{{\cal A}_{0,\, p_{1/2}}}$ and $\frac{{\cal A}_{0,\, p_{1/2}}}{{\cal A}_{0,\, p_{3/2}}}$ satisfy \Eref{A0_ratio}
with high precision.

\subsection{Heavy Neutral Atoms}

Hyperfine interaction rapidly decreases with the distance between the electron and the nucleus. Thus, to calculate the HFS constants it is necessary to know the wave function of a valence electron at small distances \cite{fese33,sobelman}.
The Coulomb field of the nucleus at such distances can be considered as unscreened and the radial wave functions of $s_{1/2}$ and $p_{1/2}$ states are proportional to each other with the coefficient $\frac{Z\alpha}{2}(1+ \frac{Z^2\alpha^2}{4})$~\cite{Hriplovich}.

According to~\Eref{HFS_nuc} the atomic factor $\eta$ is represented as:
\begin{align}\label{eta_frac}
\begin{split}
	\frac{1}{\eta} = &\frac{A_{s_{1/2}}(1,0,R)}{A_{p_{1/2}}(1,0,R)} \times\\ &\times\frac{A_{p_{1/2}}(1,0,R)-A_{p_{1/2}}(1,1,R)}{A_{s_{1/2}}(1,0,R)-A_{s_{1/2}}(1,1,R)}.
\end{split}
\end{align}
If the principal quantum numbers for both electron states are the same, then  the first fraction here is ${\cal A}_{0,\,s_{1/2}}/{\cal A}_{0,\,p_{1/2}}\approx 3$ 
up to a small BR correction~\cite{SFK78}.

The second fraction depends on the radial integrals inside the nucleus of radius $R$,
where the wave functions of $s_{1/2}$ and $p_{1/2}$ states are proportional to each other.
In this manner we obtain the leading-order term for $\eta$ as:
\begin{align}\label{eta_first_order}
        \frac{1}{\eta} = \frac{3}{4} \alpha^2 Z^2.
\end{align}
\begin{table}[htbp]
\centering
\caption{\label{tbl:eta}
The atomic factors $\eta$ for neutral Au, Tl, and Fr atoms obtained analytically from Eqs.~(\ref{eta_first_order}, \ref{eta_final}) and calculated within 
advanced correlation methods \cite{SWYG07,barzakh20,PMS19,MP95,KKDB17,chen12,KDKB18,RG20}. For the hydrogen-like ions we give the value for $n=2$, $\eta = \frac{\epsilon(2s_{1/2})}{\epsilon(2p_{1/2})}$. As discussed in the text, the dependence of $\eta$ on the principal quantum number is weak and usually can be neglected.
}
        \begin{tabular}{llll}
\hline
\\[-3mm]
                $\eta$          &Au&Tl&Fr\\
\hline
\\[-3mm]
                \Eref{eta_first_order}  &4.01&3.82&3.31\\
                \Eref{eta_final}        &3.69&3.51&3.07\\
                H-lihe ion    &4.10&3.86&3.26\\
                Neutral atom        &3.3$^{a}$&3.4(2)$^{c}$&3.1(1)$^{g}$\\
                                        &4.0(3)$^{b}$&3.1$^{d}$&3.36(5)$^{h}$\\
                                        &&2.6$^{e}$&\\
                Experiment              &--&2.84(78)$^{f}$&--\\
\hline
\end{tabular}
\\
        $^{a}$~\cite{SWYG07}; $^{b}$~\cite{barzakh20}; $^{c}$~\cite{PMS19}; $^{d}$~\cite{MP95}; $^{e}$~\cite{KKDB17}; \mbox{$^{f}$~\cite{chen12}; $^{g}$~\cite{KDKB18};
        $^{h}$~\cite{RG20}.}
\end{table}

In the next-to-leading order in $(\alpha Z)^2$ this expression modifies to:
\begin{align}\label{eta_final}
        \frac{1}{\eta} = \frac{\alpha^2 Z^2}{4} 3k \left( 1+\frac{\alpha^2 Z^2}{4}\right)^2,
\end{align}
where the coefficient $k = 1-\delta(s_{1/2})+\delta(p_{1/2})$ accounts for the BR correction and is listed in~\Tref{tbl:BR_compare}.
This expression for $\eta$ is obtained for H-like ions, but can serve as a good first approximation for neutral atoms, see~\Tref{tbl:eta}.
The dependence of the atomic factor $\eta$ on the principal quantum number $n$ can be neglected.

Note, that similar expressions can be obtained for the ratio of BR corrections:
\begin{align}\label{eta_frac2}
	\frac{1}{\eta_\mathrm{BR}} = \frac{{\cal A}_{0,\,s_{1/2}}}{{\cal A}_{0,\,p_{1/2}}} \cdot \frac{{\cal A}_{0,\,p_{1/2}}-A_{p_{1/2}}(1,0,R)}{{\cal A}_{0,\,s_{1/2}}-A_{s_{1/2}}(1,0,R)}.
\end{align}
Due to the proportionality of the radial wave functions for $s_{1/2}$ and $p_{1/2}$ states inside the nucleus  the second fractions in \Eref{eta_frac} and \Eref{eta_frac2} are the same, thus: $\frac{\eta_\mathrm{BR}}{\eta} = k.$
Table \ref{tbl:eta} shows atomic factors $\eta$ for neutral Au, Tl, and Fr atoms obtained analytically using Eqs.~(\ref{eta_first_order}, \ref{eta_final}) and calculated within different advanced correlation methods. One can see that these results are in a reasonable agreement with each other.

\section*{Conclusion}

In this work we studied the finite-nucleus-corrections to the magnetic hyperfine structure constants, namely the charge (Breit--Rosenthal) and the magnetic (Bohr--Weisskopf) corrections. These corrections break proportionality between HFS constants and nuclear $g$ factors. This effect is known as the hyperfine anomaly. Magnetic distribution inside the nucleus can drastically change from one isotope to another, while charge distribution remains almost constant. Therefore, the leading contribution to the HFA comes from magnetic corrections.


From the measured values of the differential hyperfine magnetic anomalies, one can restore their absolute values,
if the atomic factor $\eta$ -- the ratio of the Bohr-Weisskopf corrections for $s_{1/2}$ and $p_{1/2} $ states -- is calculated.
Here we studied the dependence of $\eta$ on the nuclear charge for the case of hydrogen-like ions and generalized this dependence to the case of neutral atoms.

If the atomic factor $\eta$ is calculated and the differential anomaly is measured with sufficient accuracy, then the accuracy of determining $ g $ factors of short-lived isotopes can be increased by more than an order of magnitude.
The nuclear factor $d_\mathrm {nuc}$ characterizes the valence configuration of the nucleons. For the complex nuclear configurations, which can not be accurately described by the nuclear shell model, the value of $d_\mathrm {nuc}$ can be obtained from the analysis of the experimental values of the hyperfine constants. This way one can test various nuclear models.

\section*{Acknowledgments}
Thanks are due to Anatoly Barzakh and Vladimir Shabaev for helpful discussions. 
The work was supported by the Foundation for the advancement of theoretical physics ``BASIS''  (grant \# 17-11-136).


\begin{thebibliography}{38}
\expandafter\ifx\csname natexlab\endcsname\relax\def\natexlab#1{#1}\fi
\expandafter\ifx\csname bibnamefont\endcsname\relax
  \def\bibnamefont#1{#1}\fi
\expandafter\ifx\csname bibfnamefont\endcsname\relax
  \def\bibfnamefont#1{#1}\fi
\expandafter\ifx\csname citenamefont\endcsname\relax
  \def\citenamefont#1{#1}\fi
\expandafter\ifx\csname url\endcsname\relax
  \def\url#1{\texttt{#1}}\fi
\expandafter\ifx\csname urlprefix\endcsname\relax\def\urlprefix{URL }\fi
\providecommand{\bibinfo}[2]{#2}
\providecommand{\eprint}[2][]{\url{#2}}

\bibitem[{\citenamefont{Zhang et~al.}(2015)\citenamefont{Zhang, Tandecki,
  Collister, Aubin, Behr, Gomez, Gwinner, Orozco, Pearson, Sprouse
  et~al.}}]{Z2015}
\bibinfo{author}{\bibfnamefont{J.}~\bibnamefont{Zhang}},
  \bibinfo{author}{\bibfnamefont{M.}~\bibnamefont{Tandecki}},
  \bibinfo{author}{\bibfnamefont{R.}~\bibnamefont{Collister}},
  \bibinfo{author}{\bibfnamefont{S.}~\bibnamefont{Aubin}},
  \bibinfo{author}{\bibfnamefont{J.}~\bibnamefont{Behr}},
  \bibinfo{author}{\bibfnamefont{E.}~\bibnamefont{Gomez}},
  \bibinfo{author}{\bibfnamefont{G.}~\bibnamefont{Gwinner}},
  \bibinfo{author}{\bibfnamefont{L.}~\bibnamefont{Orozco}},
  \bibinfo{author}{\bibfnamefont{M.}~\bibnamefont{Pearson}},
  \bibinfo{author}{\bibfnamefont{G.}~\bibnamefont{Sprouse}},
  \bibnamefont{et~al.}, \bibinfo{journal}{Phys. Rev. Lett.}
  \textbf{\bibinfo{volume}{115}}, \bibinfo{pages}{042501}
  (\bibinfo{year}{2015}).

\bibitem[{\citenamefont{Grossman et~al.}(1999)\citenamefont{Grossman, Orozco,
  Pearson, Simsarian, Sprouse, and Zhao}}]{G1999}
\bibinfo{author}{\bibfnamefont{J.}~\bibnamefont{Grossman}},
  \bibinfo{author}{\bibfnamefont{L.}~\bibnamefont{Orozco}},
  \bibinfo{author}{\bibfnamefont{M.}~\bibnamefont{Pearson}},
  \bibinfo{author}{\bibfnamefont{J.}~\bibnamefont{Simsarian}},
  \bibinfo{author}{\bibfnamefont{G.}~\bibnamefont{Sprouse}}, \bibnamefont{and}
  \bibinfo{author}{\bibfnamefont{W.}~\bibnamefont{Zhao}},
  \bibinfo{journal}{Phys. Rev. Lett.} \textbf{\bibinfo{volume}{83}},
  \bibinfo{pages}{935} (\bibinfo{year}{1999}).

\bibitem[{\citenamefont{Bohr and Weisskopf}(1950)}]{BW50}
\bibinfo{author}{\bibfnamefont{A.}~\bibnamefont{Bohr}} \bibnamefont{and}
  \bibinfo{author}{\bibfnamefont{V.~F.} \bibnamefont{Weisskopf}},
  \bibinfo{journal}{Phys. Rev.} \textbf{\bibinfo{volume}{77}},
  \bibinfo{pages}{94} (\bibinfo{year}{1950}).

\bibitem[{\citenamefont{Rosenthal and Breit}(1932)}]{RB32}
\bibinfo{author}{\bibfnamefont{J.~E.} \bibnamefont{Rosenthal}}
  \bibnamefont{and} \bibinfo{author}{\bibfnamefont{G.}~\bibnamefont{Breit}},
  \bibinfo{journal}{Phys. Rev.} \textbf{\bibinfo{volume}{41}},
  \bibinfo{pages}{459} (\bibinfo{year}{1932}).

\bibitem[{\citenamefont{Crawford and Schawlow}(1949)}]{CS49}
\bibinfo{author}{\bibfnamefont{M.}~\bibnamefont{Crawford}} \bibnamefont{and}
  \bibinfo{author}{\bibfnamefont{A.}~\bibnamefont{Schawlow}},
  \bibinfo{journal}{Phys. Rev.} \textbf{\bibinfo{volume}{76}},
  \bibinfo{pages}{1310} (\bibinfo{year}{1949}).

\bibitem[{\citenamefont{Persson}(2013)}]{Per13}
\bibinfo{author}{\bibfnamefont{J.~R.} \bibnamefont{Persson}},
  \bibinfo{journal}{At. Data Nucl. Data Tables} \textbf{\bibinfo{volume}{99}},
  \bibinfo{pages}{62} (\bibinfo{year}{2013}).

\bibitem[{\citenamefont{Andreyev et~al.}(2000)\citenamefont{Andreyev, Huyse,
  Van~Duppen, Weissman, Ackermann, Gerl, Hessberger, Hofmann, Kleinb{\"o}hl,
  M{\"u}nzenberg et~al.}}]{shape_coh}
\bibinfo{author}{\bibfnamefont{A.}~\bibnamefont{Andreyev}},
  \bibinfo{author}{\bibfnamefont{M.}~\bibnamefont{Huyse}},
  \bibinfo{author}{\bibfnamefont{P.}~\bibnamefont{Van~Duppen}},
  \bibinfo{author}{\bibfnamefont{L.}~\bibnamefont{Weissman}},
  \bibinfo{author}{\bibfnamefont{D.}~\bibnamefont{Ackermann}},
  \bibinfo{author}{\bibfnamefont{J.}~\bibnamefont{Gerl}},
  \bibinfo{author}{\bibfnamefont{F.}~\bibnamefont{Hessberger}},
  \bibinfo{author}{\bibfnamefont{S.}~\bibnamefont{Hofmann}},
  \bibinfo{author}{\bibfnamefont{A.}~\bibnamefont{Kleinb{\"o}hl}},
  \bibinfo{author}{\bibfnamefont{G.}~\bibnamefont{M{\"u}nzenberg}},
  \bibnamefont{et~al.}, \bibinfo{journal}{Nature}
  \textbf{\bibinfo{volume}{405}}, \bibinfo{pages}{430} (\bibinfo{year}{2000}).

\bibitem[{\citenamefont{Stroke et~al.}(1961)\citenamefont{Stroke, Blin-Stoyle,
  and Jaccarino}}]{stroke61}
\bibinfo{author}{\bibfnamefont{H.}~\bibnamefont{Stroke}},
  \bibinfo{author}{\bibfnamefont{R.~J.} \bibnamefont{Blin-Stoyle}},
  \bibnamefont{and}
  \bibinfo{author}{\bibfnamefont{V.}~\bibnamefont{Jaccarino}},
  \bibinfo{journal}{Phys. Rev.} \textbf{\bibinfo{volume}{123}},
  \bibinfo{pages}{1326} (\bibinfo{year}{1961}).

\bibitem[{\citenamefont{Pyykk{\"o} et~al.}(1973)\citenamefont{Pyykk{\"o},
  Pajanne, and Inokuti}}]{pyykko73}
\bibinfo{author}{\bibfnamefont{P.}~\bibnamefont{Pyykk{\"o}}},
  \bibinfo{author}{\bibfnamefont{E.}~\bibnamefont{Pajanne}}, \bibnamefont{and}
  \bibinfo{author}{\bibfnamefont{M.}~\bibnamefont{Inokuti}},
  \bibinfo{journal}{Int. J. Quantum Chem.} \textbf{\bibinfo{volume}{7}},
  \bibinfo{pages}{785} (\bibinfo{year}{1973}).

\bibitem[{\citenamefont{Shabaev}(1994)}]{Sha94}
\bibinfo{author}{\bibfnamefont{V.~M.} \bibnamefont{Shabaev}},
  \bibinfo{journal}{J. Phys. B} \textbf{\bibinfo{volume}{27}},
  \bibinfo{pages}{5825} (\bibinfo{year}{1994}).

\bibitem[{\citenamefont{Sushkov et~al.}(1978)\citenamefont{Sushkov, Flambaum,
  and Khriplovich}}]{SFK78}
\bibinfo{author}{\bibfnamefont{O.}~\bibnamefont{Sushkov}},
  \bibinfo{author}{\bibfnamefont{V.}~\bibnamefont{Flambaum}}, \bibnamefont{and}
  \bibinfo{author}{\bibfnamefont{I.}~\bibnamefont{Khriplovich}},
  \bibinfo{journal}{Optics and Spectroscopy} \textbf{\bibinfo{volume}{44}},
  \bibinfo{pages}{2} (\bibinfo{year}{1978}).

\bibitem[{\citenamefont{Andrae}(2000)}]{fermi_model}
\bibinfo{author}{\bibfnamefont{D.}~\bibnamefont{Andrae}},
  \bibinfo{journal}{Phys. Rep.} \textbf{\bibinfo{volume}{336}},
  \bibinfo{pages}{413 } (\bibinfo{year}{2000}), ISSN \bibinfo{issn}{0370-1573}.

\bibitem[{\citenamefont{Bohr}(1951)}]{B51}
\bibinfo{author}{\bibfnamefont{A.}~\bibnamefont{Bohr}}, \bibinfo{journal}{Phys.
  Rev.} \textbf{\bibinfo{volume}{81}}, \bibinfo{pages}{331}
  (\bibinfo{year}{1951}).

\bibitem[{\citenamefont{B{\"u}ttgenbach}(1984)}]{odd_odd_fr}
\bibinfo{author}{\bibfnamefont{S.}~\bibnamefont{B{\"u}ttgenbach}},
  \bibinfo{journal}{Hyperfine Interact.} \textbf{\bibinfo{volume}{20}},
  \bibinfo{pages}{1} (\bibinfo{year}{1984}).

\bibitem[{\citenamefont{M{\aa}rtesson-Pendrill}(1995)}]{MP95}
\bibinfo{author}{\bibfnamefont{A.-M.} \bibnamefont{M{\aa}rtesson-Pendrill}},
  \bibinfo{journal}{Phys. Rev. Lett.} \textbf{\bibinfo{volume}{74}},
  \bibinfo{pages}{2184} (\bibinfo{year}{1995}).

\bibitem[{\citenamefont{Konovalova et~al.}(2017)\citenamefont{Konovalova,
  Kozlov, Demidov, and A.E.}}]{KKDB17}
\bibinfo{author}{\bibfnamefont{E.}~\bibnamefont{Konovalova}},
  \bibinfo{author}{\bibfnamefont{M.}~\bibnamefont{Kozlov}},
  \bibinfo{author}{\bibfnamefont{Y.}~\bibnamefont{Demidov}}, \bibnamefont{and}
  \bibinfo{author}{\bibfnamefont{B.}~\bibnamefont{A.E.}},
  \bibinfo{journal}{Rad. Applic.} \textbf{\bibinfo{volume}{2}},
  \bibinfo{pages}{181} (\bibinfo{year}{2017}), ISSN \bibinfo{issn}{2466-4294}.

\bibitem[{\citenamefont{Safronova et~al.}(1999)\citenamefont{Safronova,
  Johnson, and Derevianko}}]{SJD99}
\bibinfo{author}{\bibfnamefont{M.}~\bibnamefont{Safronova}},
  \bibinfo{author}{\bibfnamefont{W.}~\bibnamefont{Johnson}}, \bibnamefont{and}
  \bibinfo{author}{\bibfnamefont{A.}~\bibnamefont{Derevianko}},
  \bibinfo{journal}{Phys. Rev. A} \textbf{\bibinfo{volume}{60}},
  \bibinfo{pages}{4476} (\bibinfo{year}{1999}).

\bibitem[{\citenamefont{Gomez et~al.}(2008)\citenamefont{Gomez, Aubin, Orozco,
  Sprouse, Iskrenova-Tchoukova, and Safronova}}]{GAOS08}
\bibinfo{author}{\bibfnamefont{E.}~\bibnamefont{Gomez}},
  \bibinfo{author}{\bibfnamefont{S.}~\bibnamefont{Aubin}},
  \bibinfo{author}{\bibfnamefont{L.~A.} \bibnamefont{Orozco}},
  \bibinfo{author}{\bibfnamefont{G.~D.} \bibnamefont{Sprouse}},
  \bibinfo{author}{\bibfnamefont{E.}~\bibnamefont{Iskrenova-Tchoukova}},
  \bibnamefont{and} \bibinfo{author}{\bibfnamefont{M.~S.}
  \bibnamefont{Safronova}}, \bibinfo{journal}{Phys. Rev. Lett.}
  \textbf{\bibinfo{volume}{100}}, \bibinfo{pages}{172502}
  (\bibinfo{year}{2008}).

\bibitem[{\citenamefont{Sahoo et~al.}(2015)\citenamefont{Sahoo, Nandy, Das, and
  Sakemi}}]{sahoo15}
\bibinfo{author}{\bibfnamefont{B.}~\bibnamefont{Sahoo}},
  \bibinfo{author}{\bibfnamefont{D.}~\bibnamefont{Nandy}},
  \bibinfo{author}{\bibfnamefont{B.}~\bibnamefont{Das}}, \bibnamefont{and}
  \bibinfo{author}{\bibfnamefont{Y.}~\bibnamefont{Sakemi}},
  \bibinfo{journal}{Phys. Rev. A} \textbf{\bibinfo{volume}{91}},
  \bibinfo{pages}{042507} (\bibinfo{year}{2015}).

\bibitem[{\citenamefont{Ginges et~al.}(2017)\citenamefont{Ginges, Volotka, and
  Fritzsche}}]{GVF17}
\bibinfo{author}{\bibfnamefont{J.}~\bibnamefont{Ginges}},
  \bibinfo{author}{\bibfnamefont{A.}~\bibnamefont{Volotka}}, \bibnamefont{and}
  \bibinfo{author}{\bibfnamefont{S.}~\bibnamefont{Fritzsche}},
  \bibinfo{journal}{Phys. Rev. A} \textbf{\bibinfo{volume}{96}},
  \bibinfo{pages}{062502} (\bibinfo{year}{2017}).

\bibitem[{\citenamefont{M{\aa}rtensson-Pendrill}(2000)}]{pendrill_fr}
\bibinfo{author}{\bibfnamefont{A.-M.} \bibnamefont{M{\aa}rtensson-Pendrill}},
  \bibinfo{journal}{Hyperfine Interact.} \textbf{\bibinfo{volume}{127}},
  \bibinfo{pages}{41} (\bibinfo{year}{2000}).

\bibitem[{\citenamefont{Johnson and Soff}(1985)}]{JS85}
\bibinfo{author}{\bibfnamefont{W.}~\bibnamefont{Johnson}} \bibnamefont{and}
  \bibinfo{author}{\bibfnamefont{G.}~\bibnamefont{Soff}},
  \bibinfo{journal}{At. Data Nucl. Data Tables}
  \textbf{\bibinfo{volume}{33}}, \bibinfo{pages}{405} (\bibinfo{year}{1985}).

\bibitem[{\citenamefont{Shabaev et~al.}(2001)\citenamefont{Shabaev, Artemyev,
  Yerokhin, Zherebtsov, and Soff}}]{shab01}
\bibinfo{author}{\bibfnamefont{V.}~\bibnamefont{Shabaev}},
  \bibinfo{author}{\bibfnamefont{A.}~\bibnamefont{Artemyev}},
  \bibinfo{author}{\bibfnamefont{V.}~\bibnamefont{Yerokhin}},
  \bibinfo{author}{\bibfnamefont{O.}~\bibnamefont{Zherebtsov}},
  \bibnamefont{and} \bibinfo{author}{\bibfnamefont{G.}~\bibnamefont{Soff}},
  \bibinfo{journal}{Phys. Rev. Lett.} \textbf{\bibinfo{volume}{86}},
  \bibinfo{pages}{3959} (\bibinfo{year}{2001}).

\bibitem[{\citenamefont{Karpeshin and Trzhaskovskaya}(2015)}]{karpeshin15}
\bibinfo{author}{\bibfnamefont{F.}~\bibnamefont{Karpeshin}} \bibnamefont{and}
  \bibinfo{author}{\bibfnamefont{M.}~\bibnamefont{Trzhaskovskaya}},
  \bibinfo{journal}{Nucl. Phys. A} \textbf{\bibinfo{volume}{941}},
  \bibinfo{pages}{66} (\bibinfo{year}{2015}).

\bibitem[{\citenamefont{Shabaev et~al.}(2006)\citenamefont{Shabaev, Glazov,
  Oreshkina, Volotka, Plunien, Kluge, and Quint}}]{shab_prl06}
\bibinfo{author}{\bibfnamefont{V.}~\bibnamefont{Shabaev}},
  \bibinfo{author}{\bibfnamefont{D.}~\bibnamefont{Glazov}},
  \bibinfo{author}{\bibfnamefont{N.}~\bibnamefont{Oreshkina}},
  \bibinfo{author}{\bibfnamefont{A.}~\bibnamefont{Volotka}},
  \bibinfo{author}{\bibfnamefont{G.}~\bibnamefont{Plunien}},
  \bibinfo{author}{\bibfnamefont{H.-J.} \bibnamefont{Kluge}}, \bibnamefont{and}
  \bibinfo{author}{\bibfnamefont{W.}~\bibnamefont{Quint}},
  \bibinfo{journal}{Phys. Rev. Lett.} \textbf{\bibinfo{volume}{96}},
  \bibinfo{pages}{253002} (\bibinfo{year}{2006}).

\bibitem[{\citenamefont{Schmidt et~al.}(2018)\citenamefont{Schmidt, Billowes,
  Bissell, Blaum, Garcia~Ruiz, Heylen, Malbrunot-Ettenauer, Neyens,
  N{\"o}rtersh{\"a}user, Plunien et~al.}}]{schmidt18}
\bibinfo{author}{\bibfnamefont{S.}~\bibnamefont{Schmidt}},
  \bibinfo{author}{\bibfnamefont{J.}~\bibnamefont{Billowes}},
  \bibinfo{author}{\bibfnamefont{M.}~\bibnamefont{Bissell}},
  \bibinfo{author}{\bibfnamefont{K.}~\bibnamefont{Blaum}},
  \bibinfo{author}{\bibfnamefont{R.}~\bibnamefont{Garcia~Ruiz}},
  \bibinfo{author}{\bibfnamefont{H.}~\bibnamefont{Heylen}},
  \bibinfo{author}{\bibfnamefont{S.}~\bibnamefont{Malbrunot-Ettenauer}},
  \bibinfo{author}{\bibfnamefont{G.}~\bibnamefont{Neyens}},
  \bibinfo{author}{\bibfnamefont{W.}~\bibnamefont{N{\"o}rtersh{\"a}user}},
  \bibinfo{author}{\bibfnamefont{G.}~\bibnamefont{Plunien}},
  \bibnamefont{et~al.}, \bibinfo{journal}{Phys. Lett. B}
  \textbf{\bibinfo{volume}{779}}, \bibinfo{pages}{324} (\bibinfo{year}{2018}).

\bibitem[{\citenamefont{Kozlov et~al.}(2015)\citenamefont{Kozlov, Porsev,
  Safronova, and Tupitsyn}}]{KPST15}
\bibinfo{author}{\bibfnamefont{M.}~\bibnamefont{Kozlov}},
  \bibinfo{author}{\bibfnamefont{S.}~\bibnamefont{Porsev}},
  \bibinfo{author}{\bibfnamefont{M.}~\bibnamefont{Safronova}},
  \bibnamefont{and} \bibinfo{author}{\bibfnamefont{I.}~\bibnamefont{Tupitsyn}},
  \bibinfo{journal}{Comput. Phys. Commun.} \textbf{\bibinfo{volume}{195}},
  \bibinfo{pages}{199} (\bibinfo{year}{2015}).

\bibitem[{\citenamefont{Bratsev et~al.}(1977)\citenamefont{Bratsev, Deyneka,
  and Tupitsyn}}]{BDT77}
\bibinfo{author}{\bibfnamefont{V.~F.} \bibnamefont{Bratsev}},
  \bibinfo{author}{\bibfnamefont{G.~B.} \bibnamefont{Deyneka}},
  \bibnamefont{and} \bibinfo{author}{\bibfnamefont{I.~I.}
  \bibnamefont{Tupitsyn}}, \bibinfo{journal}{Bull. Acad. Sci. USSR, Phys. Ser.}
  \textbf{\bibinfo{volume}{41}}, \bibinfo{pages}{173} (\bibinfo{year}{1977}).

\bibitem[{\citenamefont{Angeli and Marinova}(2013)}]{angeli13}
\bibinfo{author}{\bibfnamefont{I.}~\bibnamefont{Angeli}} \bibnamefont{and}
  \bibinfo{author}{\bibfnamefont{K.}~\bibnamefont{Marinova}},
  \bibinfo{journal}{At. Data Nucl. Data Tables} \textbf{\bibinfo{volume}{99}},
  \bibinfo{pages}{69} (\bibinfo{year}{2013}).

\bibitem[{\citenamefont{Fermi and Segr{\`e}}(1933)}]{fese33}
\bibinfo{author}{\bibfnamefont{E.}~\bibnamefont{Fermi}} \bibnamefont{and}
  \bibinfo{author}{\bibfnamefont{E.}~\bibnamefont{Segr{\`e}}},
  \bibinfo{journal}{Zeitschrift f{\"u}r Physik} \textbf{\bibinfo{volume}{82}},
  \bibinfo{pages}{729} (\bibinfo{year}{1933}).

\bibitem[{\citenamefont{Sobelman}(1977)}]{sobelman}
\bibinfo{author}{\bibfnamefont{I.~I.} \bibnamefont{Sobelman}},
  \emph{\bibinfo{title}{Introduction to the theory of Atomic Spectra}}
  (\bibinfo{publisher}{Nauka}, \bibinfo{year}{1977}).

\bibitem[{\citenamefont{Khriplovich}(1981)}]{Hriplovich}
\bibinfo{author}{\bibfnamefont{I.}~\bibnamefont{Khriplovich}},
  \emph{\bibinfo{title}{Parity nonconservation in atomic phenomena}}
  (\bibinfo{publisher}{Nauka, Moscow}, \bibinfo{year}{1981}).

\bibitem[{\citenamefont{Song et~al.}(2007)\citenamefont{Song, Wang, Ye, and
  Jiang}}]{SWYG07}
\bibinfo{author}{\bibfnamefont{S.}~\bibnamefont{Song}},
  \bibinfo{author}{\bibfnamefont{G.}~\bibnamefont{Wang}},
  \bibinfo{author}{\bibfnamefont{A.}~\bibnamefont{Ye}}, \bibnamefont{and}
  \bibinfo{author}{\bibfnamefont{G.}~\bibnamefont{Jiang}}, \bibinfo{journal}{J.
  Phys. B} \textbf{\bibinfo{volume}{40}}, \bibinfo{pages}{475}
  (\bibinfo{year}{2007}).

\bibitem[{\citenamefont{Barzakh et~al.}(2020)\citenamefont{Barzakh, Atanasov,
  Andreyev, Al~Monthery, Althubiti, Andel, Antalic, Blaum, Cocolios, Cubiss
  et~al.}}]{barzakh20}
\bibinfo{author}{\bibfnamefont{A.~E.} \bibnamefont{Barzakh}},
  \bibinfo{author}{\bibfnamefont{D.}~\bibnamefont{Atanasov}},
  \bibinfo{author}{\bibfnamefont{A.~N.} \bibnamefont{Andreyev}},
  \bibinfo{author}{\bibfnamefont{M.}~\bibnamefont{Al~Monthery}},
  \bibinfo{author}{\bibfnamefont{N.~A.} \bibnamefont{Althubiti}},
  \bibinfo{author}{\bibfnamefont{B.}~\bibnamefont{Andel}},
  \bibinfo{author}{\bibfnamefont{S.}~\bibnamefont{Antalic}},
  \bibinfo{author}{\bibfnamefont{K.}~\bibnamefont{Blaum}},
  \bibinfo{author}{\bibfnamefont{T.~E.} \bibnamefont{Cocolios}},
  \bibinfo{author}{\bibfnamefont{J.~G.} \bibnamefont{Cubiss}},
  \bibnamefont{et~al.}, \bibinfo{journal}{Phys. Rev. C}
  \textbf{\bibinfo{volume}{101}}, \bibinfo{pages}{034308}
  (\bibinfo{year}{2020}).

\bibitem[{\citenamefont{Prosnyak et~al.}(2020)\citenamefont{Prosnyak, Maison,
  and Skripnikov}}]{PMS19}
\bibinfo{author}{\bibfnamefont{S.}~\bibnamefont{Prosnyak}},
  \bibinfo{author}{\bibfnamefont{D.}~\bibnamefont{Maison}}, \bibnamefont{and}
  \bibinfo{author}{\bibfnamefont{L.}~\bibnamefont{Skripnikov}},
  \bibinfo{journal}{J. Chem. Phys.} \textbf{\bibinfo{volume}{152}},
  \bibinfo{pages}{044301} (\bibinfo{year}{2020}).

\bibitem[{\citenamefont{Chen et~al.}(2012)\citenamefont{Chen, Fan, Chen, Lin,
  Chen, Shy, and Liu}}]{chen12}
\bibinfo{author}{\bibfnamefont{T.-L.} \bibnamefont{Chen}},
  \bibinfo{author}{\bibfnamefont{I.}~\bibnamefont{Fan}},
  \bibinfo{author}{\bibfnamefont{H.-C.} \bibnamefont{Chen}},
  \bibinfo{author}{\bibfnamefont{C.-Y.} \bibnamefont{Lin}},
  \bibinfo{author}{\bibfnamefont{S.-E.} \bibnamefont{Chen}},
  \bibinfo{author}{\bibfnamefont{J.-T.} \bibnamefont{Shy}}, \bibnamefont{and}
  \bibinfo{author}{\bibfnamefont{Y.-W.} \bibnamefont{Liu}},
  \bibinfo{journal}{Phys. Rev. A} \textbf{\bibinfo{volume}{86}},
  \bibinfo{pages}{052524} (\bibinfo{year}{2012}).

\bibitem[{\citenamefont{Konovalova et~al.}(2018)\citenamefont{Konovalova,
  Demidov, Kozlov, and Barzakh}}]{KDKB18}
\bibinfo{author}{\bibfnamefont{E.}~\bibnamefont{Konovalova}},
  \bibinfo{author}{\bibfnamefont{Y.}~\bibnamefont{Demidov}},
  \bibinfo{author}{\bibfnamefont{M.}~\bibnamefont{Kozlov}}, \bibnamefont{and}
  \bibinfo{author}{\bibfnamefont{A.}~\bibnamefont{Barzakh}},
  \bibinfo{journal}{Atoms} \textbf{\bibinfo{volume}{6}}, \bibinfo{pages}{39}
  (\bibinfo{year}{2018}).

\bibitem[{\citenamefont{Roberts and Ginges}(2020)}]{RG20}
\bibinfo{author}{\bibfnamefont{B.}~\bibnamefont{Roberts}} \bibnamefont{and}
  \bibinfo{author}{\bibfnamefont{J.}~\bibnamefont{Ginges}},
  \bibinfo{journal}{arXiv preprint arXiv:2001.01907}  (\bibinfo{year}{2020}).

\end{thebibliography}
\end{document}